\documentclass[prd,
twocolumn,superscriptaddress,preprintnumbers,nofootinbib]{revtex4}
\usepackage{graphicx}
\usepackage{epsfig}
\usepackage{bm}
\usepackage{latexsym,amssymb,amsmath,amssymb,wasysym,float}

\usepackage{color}

\usepackage{scrextend}
\usepackage{mathrsfs}
\usepackage{enumitem}

\newcommand{\postscript}[2]{\setlength{\epsfxsize}{#2\hsize}
   \centerline{\epsfbox{#1}}}

\usepackage[usenames,dvipsnames]{xcolor}
\definecolor{orange}{cmyk}{0,0.5,1,0}
\definecolor{rossoCP3}{cmyk}{0,.88,.77,.40}
\definecolor{graa}{rgb}{0.8,0.8,0.8}
\definecolor{blaa}{rgb}{0.2,0.2,0.6}

\begin{document}

\preprint{MPP-2024-219}
\preprint{LMU-ASC 19/24}

\title{\color{rossoCP3} Cosmological Constraints on Dark Neutrino Towers}

\author{Luis A. Anchordoqui}

\affiliation{Department of Physics and Astronomy,\\  Lehman College, City University of
  New York, NY 10468, USA
}

\affiliation{Department of Physics,\\
 Graduate Center,  City University of
  New York,  NY 10016, USA
}

\affiliation{Department of Astrophysics,
 American Museum of Natural History, NY
 10024, USA
}

\author{Ignatios Antoniadis}

\affiliation{High Energy Physics Research Unit, Faculty of Science, Chulalongkorn University, Bangkok 1030, Thailand}

\affiliation{Laboratoire de Physique Th\'eorique et Hautes \'Energies
  - LPTHE \\
Sorbonne Universit\'e, CNRS, 4 Place Jussieu, 75005 Paris, France
}

\author{Dieter\nolinebreak~L\"ust}

\affiliation{Max--Planck--Institut f\"ur Physik,  
 Werner--Heisenberg--Institut,
80805 M\"unchen, Germany
}

\affiliation{Arnold Sommerfeld Center for Theoretical Physics, 
Ludwig-Maximilians-Universit\"at M\"unchen,
80333 M\"unchen, Germany
}

\author{Karem Pe\~nal\'o Castillo}
\affiliation{Department of Physics and Astronomy,\\  Lehman College, City University of
  New York, NY 10468, USA
}

\begin{abstract}
  \noindent We reexamine a dynamical dark matter model with
  Kaluza-Klein (KK)
  towers of gravitons and neutrinos fitting together in the dark
  dimension. We show that even though gravitational decays of neutrino
  KK towers have little impact in cosmology the weak decay channel
  could have significant cosmological effects. Taking conservative
  upper bounds on the dark matter decay rate into two photons before
  recombination and on the number of effective extra neutrino species
  $\Delta N_{\rm eff}$ we derive constraints on the conversion rate from active to sterile species
  despite the dependence of the mixing angle on the KK mode mass. We
  also provide counterarguments to a recent claim suggesting that the bounds on
$\Delta N_{\rm eff}$ rule out micron-sized extra dimensions.
\end{abstract}
\date{November 2024}
\maketitle

\section{Introduction}

The mechanism behind neutrino masses is now in limbo. One interesting
proposal envisions the right-handed neutrinos as five-dimensional (5D)
bulk states, with Yukawa couplings to the left-handed lepton and Higgs doublets that are localized states on the
Standard Model (SM)
brane~\cite{Dienes:1998sb,Arkani-Hamed:1998wuz,Dvali:1999cn}. The
neutrino Yukawa couplings become tiny because of a volume
suppression, yielding naturally light Dirac neutrinos. As a
by-product of this type of construct, a neutrino tower of Kaluza-Klein (KK) modes arises with masses proportional to the inverse
radius $R_\perp$ of the fifth dimension.

Very recently, cosmological constraints on neutrino KK towers were
reconsidered~\cite{McKeen:2024fdl}, with a deceptive conclusion that current bounds on the
effective number of noninteracting relativistic species from
measurements of the cosmic microwave background (CMB)~\cite{Planck:2018vyg} generically rule out micron-sized extra
dimensions. In this paper we show that the limits inferred
in~\cite{McKeen:2024fdl} are not generic but rather model
dependent. A major shortcoming of the analysis carried out
in~\cite{McKeen:2024fdl} is that the assumed cosmic evolution of the
KK tower does not allow for intra-tower neutrino decays, {\it viz.} decays
of a given KK mode in the tower into final states that include other,
lighter KK excitations. Such a ``dark-to-dark'' decay process usually
dominates the cosmological evolution of the tower and could be
regarded as the telltale
signature of the dynamical dark matter framework~\cite{Dienes:2011ja}. We also show that
the limits
on $R_\perp$ inferred in~\cite{McKeen:2024fdl}  do not particularly apply to the
dark dimension scenario~\cite{Montero:2022prj}.

The layout of the paper is as follows. We begin in Sec.~\ref{sec:2} by
summarizing phenomenological aspects of the dark dimension
scenario. Along the way, we also provide a brief review of the
cosmological evolution of dark
gravitons and their interplay
with dark neutrino towers~\cite{Gonzalo:2022jac}. In Sec.~\ref{sec:3} we reexamine
cosmological constraints
on weak decays of bulk sterile neutrinos into active neutrinos and
photons. The paper wraps up in Sec.~\ref{sec:4} with some conclusions.

\section{Dynamical Dark Matter with Towers of Gravitons and Neutrinos}
\label{sec:2}

The dark dimension is a five-dimensional (5D) setup that has a compact
space with characteristic length-scale in the micron range. Most notably, this
setup provides an economic
explanation of the
cosmological hierarchy problem, because the anti-de Sitter
distance conjecture in de Sitter
space~\cite{Lust:2019zwm} connects the size of the compact space $R_\perp$ to the dark energy scale $\Lambda^{1/4}$
via $R_\perp \sim \lambda \Lambda^{1/4}$, where
$\Lambda \sim 10^{-120} M_p^4$ is the cosmological constant, $M_p$ the
reduced Planck mass, and the proportionality factor is estimated to
be within the range $10^{-1} < \lambda <
10^{-4}$~\cite{Montero:2022prj}. The KK  tower of the dark dimension
opens up at the mass scale $m_{\rm KK} \sim 1/R_\perp$. The species
scale $M_*$ where gravity becomes strong is linked to $m_{\rm KK}$ via
$M_* \sim m_{\rm KK}^{1/3} \
M_p^{2/3}$~\cite{Dvali:2007hz,Dvali:2007wp}. In our calculations we
adopt $m_{\rm KK} \sim 0.1~{\rm eV}$ which implies $M_* \sim 10^9~{\rm GeV}$.

The dark dimension assembles a colosseum for dark matter (DM) 
contenders. Primordial black holes with Schwarzschild radius smaller
than a micron provide one interesting DM candidate~\cite{Anchordoqui:2022txe,Anchordoqui:2022tgp,Anchordoqui:2024akj,Anchordoqui:2024dxu,Anchordoqui:2024jkn,Anchordoqui:2024tdj}. Massive spin-2 KK
excitations of the graviton take the place of a second compelling
possibility~\cite{Gonzalo:2022jac}. Herein we focus attention on the
second scenario whereby DM production proceeds through the coupling of SM fields with
the 5D graviton. The cosmological chronicle begins with the matter
fields in equilibrium at initial temperature $T_{\rm in} \sim~{\rm
  GeV}$~\cite{Gonzalo:2022jac}. These are localized recurrences in the compact space, which manifest as a
tower of equally spaced dark gravitons, indexed by an integer $l$, and with mass  $m_l = l \ m_{\rm KK}$. The production of graviton mode of mass $m_l$
is dominantly
at temperatures $T_l \sim m_l$. Thereby, the produced dark gravitons are initially
in the GeV mass range. 

A point worth noting at this juncture is that the cosmological
overproduction of bulk graviton modes implies that in 
models with $n$ large or mesoscopic extra dimensions  the maximum temperatures must be less than
\begin{equation}
  T_\star \sim 10^{3+(6n-15)/(n+2)} \left(\frac{M_*}{10^9~{\rm GeV}}\right)~{\rm GeV}
\end{equation}
which is an upper limit on the ``normalcy'' temperature at which the
universe must be free of bulk modes~\cite{Arkani-Hamed:1998sfv}. Note that for $n=1$, the assumed $T_{\rm
  in} \sim 1~{\rm GeV}$ saturates the normalcy temperature.

The cosmic evolution of the dark sector is mostly driven by dark-to-dark
decay processes, which regulate the decay of KK gravitons within the
dark tower, realizing a particular version of the dynamical dark
matter framework~\cite{Dienes:2011ja}. In the absence of isometries
in the dark dimension, which is the common expectation, the KK momentum of the
dark tower is not conserved~\cite{Mohapatra:2003ah}. This means that a dark graviton of KK
quantum $n$ can decay to two other ones, with quantum numbers $n_1$
and $n_2$. If the KK quantum violation can go
up to $\delta n$, the number of available channels is roughly $l \,
\delta n$. In addition, because the decay is almost at threshold, the phase space
factor is roughly the velocity of decay products,
  $v_{\rm r.m.s.} \sim \sqrt{m_{\rm KK} \ \delta n /m_l}$. All in all, the total
  decay width of graviton $G_l$ is found to be, 
\begin{eqnarray}
  \Gamma^{G_l}_{\rm tot} & \sim & \sum_{l'<l} \ \ \sum_{0<l''<l-l'}
                              \Gamma^{G_l}_{G_{l'} G_{l''}} \nonumber \\
  & \sim &
  \beta^2 \frac{m_l^3}{M_p^2} \times \frac{m_l}{m_{\rm KK}} \ \delta
  n \times \sqrt{\frac{m_{\rm KK} \delta n}{m_l} } \nonumber \\
  & \sim & \beta^2 \
    \delta n^{3/2} \frac{m_l^{7/2}}{M_p^2 m_{\rm KK}^{1/2}} \,,
\label{Gtot}
\end{eqnarray}   
 where $\beta$ is parameter that controls the strength of the intra-tower decay amplitudes which correlates with the amplitudes on inhomogeneities in the dark dimension~\cite{Gonzalo:2022jac}. We further follow~\cite{Gonzalo:2022jac} to estimate the time
evolution of the dark matter mass and
assume that for times larger than $1/\Gamma^{G_l}_{\rm tot}$ dark matter
which is
heavier than the corresponding $m_l$ has already decayed, yielding
\begin{equation}
  m_l \sim \left(\frac{M_p^4 \ m_{\rm KK}}{\beta^4 \ \delta n^3}\right)^{1/7} t^{-2/7} \,,
\label{mevol}
\end{equation}
where $t$ indicates the time elapsed since the big bang.

The dark dimension scenario also
provides visible signals on the brane, e.g., the decay of relic
graviton KK states to photons. The partial decay width of such a
process is estimated to be
\begin{equation}
  \Gamma^{G_{l}}_{\gamma \gamma} = \frac{{\tilde \lambda}^2 \ m^3_{\rm KK} \ l^3}{80 \pi
    M_p^2} \,,
\label{Ggammagamma}
\end{equation}
where the parameter $\tilde \lambda$ measures the value of the dark graviton wave
function at the SM brane and is expected to be ${\cal O}(1)$~\cite{Han:1998sg,Hall:1999mk}.

Hitherto, dark dimension cosmology can be characterized by a set of three
parameters $\{\beta, \delta n, \tilde \lambda\}$, which must be
constrained by experiment. In particular, CMB measurements are sensitive to processes that change 
the photon-baryon fluid between today and the last
scattering surface. To be more specific,
throughout the CMB epoch during the emission of the relic at redshift $z_{\rm CMB} \sim 1100$, matter dominates the energy
density of the Universe. The number density of baryons is mostly
composed of neutral hydrogen atoms (${\rm H}_{\rm I}$), together with a smaller Helium
(He) component, $x_{\rm He} = n_{\rm He}/n_{{\rm H}_{\rm I}} \simeq 1/13$, and a
small percentage of free protons and electrons, $x_e = n_e /n_{{\rm
    H}_{\rm I}} =
n_p /n_{{\rm H}_{\rm I}}$, which varies from about 20\% at $z_{\rm CMB}$ to roughly
$2 \times 10^{-4}$ at $z \sim 20$~\cite{Ali-Haimoud:2010tlj}. Most notably, the damping of the CMB power
spectrum is driven by the integrated optical depth along the line of
sight. Now, dark graviton towers decaying into photons within the
redshift range $20 \alt z \alt 1100$ would bring in {\it exotic}
energy that could ionize ${\rm H}_{\rm I}$ and He. This in turn would increase the optical depth to
recombination inducing a stronger damping of the CMB power spectrum on small
scales. The non-observation of such a suppression in the spectrum
places an upper limit on the energy delivered by DM decays
into $\gamma \gamma$~\cite{Slatyer:2016qyl}, providing a direct constraint on $\tilde \lambda$~\cite{Law-Smith:2023czn}. 

In addition, we have seen that dark matter decay gives the daughter particles a
velocity kick. Self-gravitating dark-matter halos that have a virial
velocity smaller than this velocity kick may be disrupted by these
particle decays. Combined cosmological
zoom-in simulations of decaying dark matter with a  model of the Milky
Way satellite population rule out
non-relativistic kick speeds $\agt 10^{-4}$ for a dark matter
lifetime $\tau_{\rm DM} \alt 
29~{\rm Gyr}$ at 95\%CL~\cite{DES:2022doi}. However,  N-body simulations of
isolated dark-matter halos seem to indicate that if $\tau_{\rm DM} \agt
60~{\rm Gyr}$ and the kick
speed $\alt 10^{-2}$ then the halos are essentially
unchanged~\cite{Peter:2010jy}. Herein, we adopt the 95\%CL bound on
today's dark matter kick velocity $\leq 2.2 \times 10^{-4}$ derived in~\cite{Obied:2023clp}, which directly constrain the $\beta$ and $\delta n$ parameters.

We adopt as benchmark the following parameter set $\{\beta \sim 500,\
  \delta n \sim 0.2,\ \tilde{\lambda} \sim 0.5 \}$ which is consistent
  with all available data; see Fig.~4 in~\cite{Obied:2023clp}. Substituting these
  figures into (\ref{Gtot}) and (\ref{Ggammagamma}) we obtain the following parametrization of the
  decay rates as a function of $m_l$,
\begin{equation}
  \Gamma^{G_l}_{\rm tot} 
 \sim  5 \times
    10^{-4} \  \left(\frac{m_l}{{\rm
          GeV}}\right)^{7/2}~{\rm s}^{-1}
\label{Gtotnum}
  \end{equation}
and
\begin{equation}
  \Gamma^{G_{l}}_{\gamma \gamma} \sim 2 \times 10^{-16} \
  \left(\frac{m_l}{{\rm GeV}}\right)^{3}~{\rm s}^{-1} \, ,
\label{Ggamma}
\end{equation}
respectively. We can also conclude substituting the parameters $\beta$ and
$\delta n$ into (\ref{mevol}) that at present the DM has a mass
$m_{l,{\rm today}} \sim 78~{\rm keV}$, where we have taken an age of
the universe of 13.8~Gyr. 

Thus far, we have assumed that DM is composed entirely of dark
gravitons, but as it was already put forward in~\cite{Gonzalo:2022jac}
the composition of the dynamical
dark matter ensemble could call for other light modes in the bulk, which lead to new decay
channels and additional DM components. To be specific, we do expect dark
fermions to propagate in the bulk playing the role of right-handed
neutrinos. Generally speaking, one would expect that the existence of
other KK species would not affect the total abundance or cosmological
evolution of the DM mass as
the KK modes would just distribute among each other, as they decay from one to
another. For example, the $l'$-summed width of the decay $\nu_l \to
\nu_{l'} G_{l -l'}$ derived in~\cite{Abazajian:2000hw},
\begin{equation}
\sum_{l'} \Gamma^{\nu_l}_{\nu_{l'} G_{l-l'}} = \frac{ m_l^4 \ R_\perp}{6 \pi
  \ M_p^2} \sim 10^{-4} \ \left(\frac{m_l}{1~{\rm GeV}}\right)^{4}~{\rm s}^{-1} \, ,
\end{equation}
is comparable to the total decay width of the graviton (\ref{Gtotnum}),
but somewhat smaller reflecting the tiny violation 
of the KK momentum conservation, i.e. $\delta n \sim 0.2$. However, for the particular case
of dark neutrino towers we should exercise some caution, because these
towers do not only couple gravitationally to the SM
sector but also via Fermi's
weak interaction. It is this weak decay channel that we now turn to study.

\section{Constraints on bulk neutrinos}
\label{sec:3}

Consider three 5D Dirac
fermions  $\Psi_\alpha$, which are singlets under the SM gauge
symmetries and interact in our brane with the three active left-handed
neutrinos in a way that conserves lepton number.  The  $S^1/\mathbb{Z}_2$ symmetry in the dark dimension coordinate $y$
 contains $y$ to $-y$, which acts as chirality ($\gamma_5$) on
spinors. Then, in the Weyl basis each Dirac field can be decomposed into two two-component spinors $\Psi_\alpha \equiv (\psi_{\alpha L},\psi_{\alpha R})^T$.

Neutrino masses
arise in 5D bulk-brane interactions of the form
\begin{equation} 
  \mathscr{L}  \supset h_{ij} \ \overline L_i \ \tilde{H} \ \psi_{jR}(y=0) \,,
\end{equation}
where $\tilde{H} = -i\sigma_{2}H^{*}$, $L_i$ denotes the lepton
doublets (localized on the SM brane), $\psi_{jR}$ stands for the three bulk (right-handed) $R$-neutrinos
evaluated at the position of the SM brane, $y=0$ in the
dark dimension, and $h_{ij}$ are coupling
constants. This gives a coupling with the
$L$-neutrinos of the form $\langle H \rangle \  \overline{\nu}_{L_i} \
\psi_{jR} (y=0)$, where $\langle H \rangle = 175~{\rm GeV}$ is the Higgs vacuum
expectation value. Expanding $\psi_{jR}$ into modes canonically normalized leads for each of them to a Yukawa $3 \times 3$ matrix suppressed by the square root of the volume of the bulk
$\sqrt{\pi R_\perp M_s}$, i.e.,
\begin{equation}
Y_{ij}= \frac{h_{ij}}{\sqrt{\pi R_\perp M_s}} \sim h_{ij} \frac{M_s}{M_p} \,,
\end{equation}
where $M_s \lesssim M_*$ is the string scale, and where
in the second rendition we have dropped factors of $\pi$'s and of the string coupling.

Data analyses from short- and long-baseline neutrino oscillation
experiments, together with observations of neutrinos produced by
cosmic rays collisions in the atmosphere and by nuclear fusion
reactions in the Sun, provide the most sensitive insights to determine
the extremely small mass-squared differences. Neutrino oscillation data can
be well-fitted in terms of two nonzero differences $\Delta
m^2_{ij}=m^2_i-m^2_j$ between the squares of the masses of
the three ($i=1,2,3$) neutrino mass eigenvalues $m_i$, yielding $\Delta m_{\rm
  SOL}^2 = \Delta m^2_{21} = (7.53 \pm 0.18) \times 10^{-5}~{\rm
  eV}^2$  and $\Delta m^2_{\rm ATM} = |\Delta m_{31}^2| \simeq \Delta m^2_{32} = 2.453 \pm 0.033) \times 10^{-3}~{\rm
  eV}^2$ or $\Delta m^2_{32} = -2.536 \pm 0.034) \times 10^{-3}~{\rm
  eV}^2$~\cite{ParticleDataGroup:2024cfk}. A straightforward
calculation shows that to
obtain the correct order of magnitude of neutrino masses the coupling
$h_{ij}$ should be ${\cal O} (10^{-4})$  for our fiducial value $M_s \sim 10^9~{\rm GeV}$~\cite{Anchordoqui:2022svl}. 

Now, light modes in the bulk contribute to the effective number
of relativistic neutrino-like
species $N_{\rm eff}$~\cite{Steigman:1977kc} and are bounded by experiment~\cite{Planck:2018vyg}. Using conservation of entropy, fully
thermalized relics with $g_*$ degrees of freedom contribute
\begin{equation}
  \Delta N_{\rm eff} = g_* \left(\frac{43}{ 4g_s}\right)^{4/3} \left
    \{ \begin{array}{ll} 4/7 & {\rm for  \ bosons}\\ 1/2 & {\rm for \
                                                           fermions} \end{array}
               \right.                                        \,,
\end{equation}
where $g_s$ denotes the effective 
degrees of freedom for the entropy of the other thermalized
relativistic species that are present when they decouple~\cite{Anchordoqui:2011nh}. The 5D
graviton has 5 helicities, but the spin-1 helicities do not have zero
modes, because we assume the compactification has
$S^1/\mathbb{Z}_2$ symmetry and so the $\pm 1$ helicities are
projected out. The spin-0 is the
radion and
the spin-2 helicities form the massless (zero mode) graviton. This means
that for the 5D graviton, $g_*=3$. The (bulk) left-handed neutrinos are odd, but the right-handed neutrinos are even and so each
counts as a Weyl neutrino, for a total $g_* =2 \times 3$. Assuming that the
dark sector decouples from the SM sector before the electroweak phase
transition we have $g_s = 106.75$. This gives $\Delta N_{\rm eff} =
0.22$, in agreement with CMB observations~\cite{Planck:2018vyg}. Actually, $\Delta N_{\rm eff} = 0.22$ roughly saturates the value from
CMB observations.  

The KK states of the neutrino towers are non-relativistic during the CMB
formation epoch.  Indeed, in the presence of bulk masses~\cite{Lukas:2000wn,Lukas:2000rg},
the mixing of the first KK modes to active neutrinos can be
suppressed~\cite{Carena:2017qhd,Anchordoqui:2023wkm}.\footnote{We note
  in passing that bulk neutrino masses relax the constraints derived
  in~\cite{Machado:2011jt,Forero:2022skg} from neutrino oscillation experiments; for details, see~\cite{Anchordoqui:2023wkm}.} Thus, these KK states do not directly contribute to $\Delta N_{\rm eff}$. However,
these dark neutrinos can decay into SM fields (including neutrinos and
photons) via weak interactions through their small active admixture. The
active neutrinos that are produced through these decays after neutrino
decoupling at $T_\nu \sim {\rm MeV}$ but before the CMB epoch contribute
to $\Delta N_{\rm eff}$.

Next, in line with our stated plan, we estimate the unsolicited contributions to
$\Delta N_{\rm eff}$ from KK decays mediated by Fermi weak interaction dressed by an
active-sterile mixing angle. The partial decay width of the process $\nu_l \to \nu \nu \nu$ is found to
be~\cite{Abazajian:2000hw}
\begin{equation}
  \Gamma^{\nu_l}_{\nu \nu \nu}  =  \frac{G_F^2 \ m_l^5 \ \sin^2 \theta_l}{192
                                \pi^3} \sim \frac{\zeta^2 \ m_\nu^2 \
                                m_l^3}{\tau_\mu m_\mu^5} \,,
\end{equation}  
where $\theta_l \sim  \zeta m_\nu/m_l$ is  the vacuum mixing angle between an
active neutrino of mass  $m_\nu \sim 0.1~{\rm eV}$ and a KK mode with mass
$m_l$~\cite{Dodelson:1993je}, $\zeta$ parametrizes our ignorance of KK decays and neutrino mixing  in
the dark dimension, and 
\begin{equation}
G_F = \sqrt{\frac{192 \ \pi^3}{m_\mu^5 \ \tau_\mu \ (1 + \Delta q)}}  
\end{equation} 
is the Fermi constant, with $m_\mu$ and $\tau_\mu$ the muon mass and its
lifetime, respectively, and where $\Delta q \ll 1$ parametrizes the
effect of radiative corrections~\cite{MuLan:2010shf}. The partial
width of the radiative decay  $\nu_l \to\gamma \nu$ is smaller by a
factor of $27 \alpha/(8\pi)$, with $\alpha$ the fine structure
constant~\cite{Barger:1995ty}.

Now, by demanding that
$\Gamma^{\nu_l}_{\gamma \nu} \alt \Gamma^{G_l}_{\gamma \gamma}$ we
find $\zeta \alt 0.01$. Indeed, for $\zeta \sim 0.01$ we obtain
\begin{equation}
  \Gamma^{\nu_l}_{\gamma \nu} \sim 2.5 \times 10^{-16} \left(\frac{m_l}{{\rm GeV}}\right)^{3}~{\rm s}^{-1} \,,
\label{fin1}
\end{equation}
and
\begin{equation}
  \Gamma^{\nu_l}_{\nu \nu \nu}   \sim 3\times  10^{-14} \left(\frac{m_l}{{\rm GeV}}\right)^{3}~{\rm s}^{-1}  \, .
\label{fin2}
\end{equation} 
Note that the partial decay widths (\ref{fin1}) and (\ref{fin2}) are much smaller than the
inverse of the age of the universe at CMB ($t_{\rm CMB}^{-1} \sim
10^{-13}~{\rm s}^{-1}$), and so there is no contribution to $\Delta N_{\rm eff}$ from KK
weak decays. Moreover, by construction, the partial decay width of
$\nu_l \to \gamma \nu$ is comparable to the size of the width derived
in~\cite{Law-Smith:2023czn} for $G_l \to \gamma \gamma$, ensuring that for $z \agt 20$, the KK weak decay would
not inject enough energy to perturb the abundances of ${\rm H}_{\rm I}$ and He. We conclude that nowadays micron-sized extra dimensions are
consistent with observations.

\section{Conclusions}
\label{sec:4}

We have reexamined the dynamical dark matter model with KK towers of
gravitons and neutrinos proposed in~\cite{Gonzalo:2022jac}. More
specifically, we investigated the impact of weak decays from dark
neutrino towers in cosmology. Our results are encapsulated in Fig.~\ref{fig:1} and can be
summarized as follows:
\begin{figure}[htb]
  \postscript{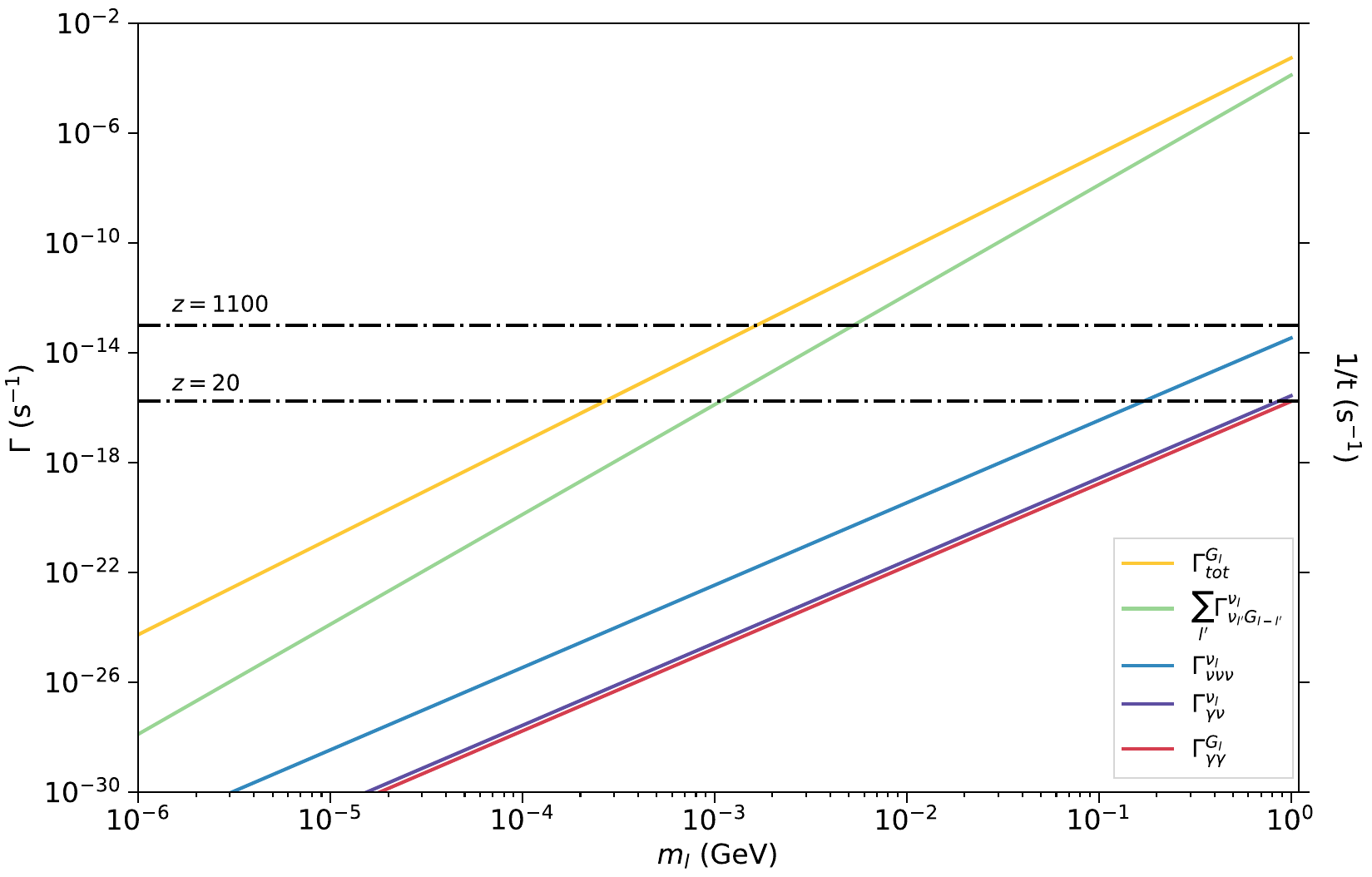}{0.9}
  \caption{Partial decay widths of various processes governing
    the cosmic evolution of the dynamical dark matter model with towers
    of gravitons and neutrinos proposed
    in~\cite{Gonzalo:2022jac}. For comparison, the horizontal   lines
    indicate the inverse of the age of the universe at $z=20$ and
    $z= 1100$, characterizing the reionization and CMB epochs, respectively.  \label{fig:1}}
\end{figure}

\begin{itemize}[noitemsep,topsep=0pt]
\item A thorough analysis of data from the {\it Planck} mission
  presented in~\cite{Law-Smith:2023czn}  provides strict constraints on the partial decay width
  $\Gamma^{G_{l}}_{\gamma \gamma}$ not to perturb the reionization
  process of the Universe.  Radiative decays $\nu_l \to \gamma \nu$,
  going via loop diagrams, can also discompose the reionization
  process. By demanding that
  $\Gamma^{\nu_{l}}_{\gamma \nu} \alt \Gamma^{G_{l}}_{\gamma \gamma} $
  we have placed a bound on the vacuum sterile mixing, which is driven by the
  parameter $\zeta$. We have shown that $\zeta \alt 0.01$ ensures that KK weak decays
  would not inject enough energy to perturb the abundances of
  ${\rm H}_{\rm I}$ and He at $z\agt 20$.
As can be seen in Fig.~\ref{fig:1} the decay widths of
$\Gamma^{\nu_{l}}_{\gamma \nu}$ and  $\Gamma^{G_{l}}_{\gamma \gamma}$
are always smaller than the inverse of the age of the universe during
reionization $t_{\rm reio}^{-1} \sim 2 \times 10^{-16}~{\rm s}^{-1}$ despite the dependence of the mixing angle on the KK mode mass.  
\item As can be deduced from Fig.~\ref{fig:1} the upper bound on the
  parameter $\zeta
  \alt 0.01$ guarantees that decay widths of processes with active
  neutrinos in the final state are all smaller than $t^{-1}_{\rm
    CMB}$. This implies that the contribution to $\Delta N_{\rm eff}$
  from weak decays of KK towers is negligible. Thus, the DM model
  proposed in~\cite{Law-Smith:2023czn} is consistent with upper bounds
  on $\Delta N_{\rm eff}$ from {\rm Planck} data.
\item In vacuum and with a seesaw model, $m_{\nu} \ll m_l$, the
  active-sterile mixing
  angle is generally approximated by $\theta_l \sim m_\nu/m_l$. The
  additional suppression factor which we have bounded to be $\zeta \alt 0.01$
  is consistent with the estimates obtained in 4D sterile neutrino models; see
  e.g.~\cite{Dodelson:1993je,Dolgov:2000ew}.

\item The cosmological evolution of the KK towers is dominated by dark-to-dark decay
  processes. Actually, the intra-KK decays in the bulk require a
  spontaneous breakdown of the translational invariance in the compact
  space such that the 5D momenta are not conserved. The
  level of violation of KK momentum conservation is very small: $\delta n \sim 0.2$. This is visible in
  Fig.~\ref{fig:1} through a comparison of the total decay width
  $\Gamma_{\rm tot}^{G_l}$ with 
  $\sum_{l'} \Gamma^{\nu_l}_{\nu_{l'} G_{l-l'}}$.
\item Alternatively, we could postulate that the violation of 5D
  momentum conservation is large
  such that KK right-handed-neutrino modes would be able decay
  fast to the zero mode, before neutrino decoupling. If this were the case, there would be
  no cosmological effects. Of course, in such a scenario DM cannot be
  realized by KK gravitons. DM can instead be due to 5D primordial
 black
 holes~\cite{Anchordoqui:2022txe,Anchordoqui:2022tgp,Anchordoqui:2024akj,Anchordoqui:2024dxu,Anchordoqui:2024jkn,Anchordoqui:2024tdj}
 or/and a fuzzy radion~\cite{Anchordoqui:2023tln}. 
\item One major assumption of the dynamical dark matter model proposed
  in~\cite{Gonzalo:2022jac}  is that the SM brane is in a thermal state at $T_{\rm in}$ and the KK modes remain
  essentially unexcited. As the SM brane
 begins cooling off, its universal coupling to bulk fields gives
 inevitable KK production.  Now, different pieces of experimental data point to an initial
  temperature just above the QCD phase transition. As noted in~\cite{Gonzalo:2022jac}, such a happenstance may actually be encrypted in
  \begin{equation}
T_{\rm in} \sim \left(\Lambda/M_p^4\right)^{1/6} M_p \sim 1~{\rm GeV} \, ,
\end{equation}
suggesting a possible UV/IR mixing at the interplay between dark energy and
$T_{\rm in}$. An investigation along these lines is obviously important to be done.
 
\end{itemize}

\section*{Acknowledgements}

The work of L.A.A. and K.P.C. is supported by the U.S. National Science
Foundation (NSF Grant PHY-2412679). I.A. is supported by the Second
Century Fund (C2F), Chulalongkorn University.  The work of D.L. is supported by the Origins
Excellence Cluster and by the German-Israel-Project (DIP) on Holography and the Swampland.

\end{document}